\pdfoutput=1

\documentclass[twocolumn]{aastex631}
\usepackage{amsmath}
\usepackage{microtype}
\usepackage{graphics}
\usepackage{subfigure}
\usepackage{float}
\usepackage{color}
\usepackage{tikz}
\usetikzlibrary{decorations.pathreplacing}
\usepackage{pgffor}
\usepackage{ulem}
\usepackage{url}
\shorttitle{VL in CG Gravity}
\shortauthors{Su et al.}
\graphicspath{{./}{figures/}}

\begin{document}

\title{Void Lensing in Cubic Galileon Gravity}

\author{Chen Su}
\affiliation{Shanghai Astronomical Observatory (SHAO), Nandan Road 80, Shanghai 200030, China}
\affiliation{University of Chinese Academy of Sciences, Beijing 100049, China}
\author{Huanyuan Shan}
\affiliation{Shanghai Astronomical Observatory (SHAO), Nandan Road 80, Shanghai 200030, China}
\affiliation{University of Chinese Academy of Sciences, Beijing 100049, China}
\author{Jiajun Zhang}
\affiliation{Shanghai Astronomical Observatory (SHAO), Nandan Road 80, Shanghai 200030, China}
\affiliation{University of Chinese Academy of Sciences, Beijing 100049, China}
\author{Cheng Zhao}
\affiliation{National Astronomy Observatories, Chinese Academy of Science, Beijing, 100101, China \\}
\affiliation{Laboratory of Astrophysics, \'Ecole Polytechnique F\'ed\'erale de Lausanne (EPFL), Observatoire de Sauverny, CH-1290 Versoix, Switzerland \\}
\author{Jiaxi Yu}
\affiliation{Laboratory of Astrophysics, \'Ecole Polytechnique F\'ed\'erale de Lausanne (EPFL), Observatoire de Sauverny, CH-1290 Versoix, Switzerland \\}
\author{Qiao Wang}
\affiliation{National Astronomy Observatories, Chinese Academy of Science, Beijing, 100101, China \\}
\affiliation{University of Chinese Academy of Sciences, Beijing 100049, China}
\author{Linfeng Xiao}
\affiliation{Department of Physics and Astronomy, Sejong University, Seoul, 143-747, Republic of Korea \\}
\author{Xiangkun Liu}
\affiliation{South-Western Institute for Astronomy Research, Yunnan University, Kunming 650500, China \\}
\author{An Zhao}
\affiliation{South-Western Institute for Astronomy Research, Yunnan University, Kunming 650500, China \\}
\correspondingauthor{Chen Su}
\email{suchen@shao.ac.cn}
\correspondingauthor{Huanyuan Shan}
\email{hyshan@shao.ac.cn}
\correspondingauthor{Jiajun Zhang}
\email{jjzhang@shao.ac.cn}



\begin{abstract}
Weak lensing studies via cosmic voids are a promising probe of Modified Gravity (MG). Excess surface mass density (ESD) is widely used as a lensing statistic in weak lensing research. In this paper, we use the ray-tracing method to study the ESD around voids in simulations based on Cubic Galileon (CG) gravity. With the compilation of N-body simulation and ray-tracing method, changes in structure formation and deflection angle resulting from MG can both be considered, making the extraction of lensing signals more realistic. We find good agreements between the measurement and theoretical prediction of ESD for CG gravity. Meanwhile, the lensing signals are much less affected by the change of the deflection angle than the change of the structure formation, indicating a good approximation of regarding ESD (statistics) as the projection of 3D dark matter density field. Finally, we demonstrate that it is impossible to distinguish CG and General Relativity in our simulation, however, in the next-generation survey, thanks to the large survey area and the increased galaxy number density, detecting the differences between these two models is possible. The methodology employed in this paper that combines N-body simulation and ray-tracing method can be a robust way to measure the lensing signals from simulations based on the MGs, and especially on that which significantly modifies the deflection angle.
\end{abstract}

\keywords{weak gravitational lensing; modified gravity; cosmic void; cosmology}


\section{Introduction} \label{sec:intro}

The discovery of the late-time acceleration of the Universe brings us a new view to the components in the universe. In the framework of General Relativity (GR), approximately 70\% of the total energy in our universe is occupied by dark energy, a component that could be mimicked by a cosmological constant, $\Lambda$, but the exact nature of which is unclear now. On the other hand, the expanded universe may suggest the need for new physics beyond the framework of Einstein's theory, which motivates the study of Modified Gravity (MG). Without the cosmological constant, these theories propose several alternatives to GR to explain the accelerating Universe.

Cubic Galileon (CG) theory, a special (and simple) case of Horndeski theory \citep{1974IJTP...10..363H, 2019RPPh...82h6901K}, is one of the scalar-tensor theories that allows modification of GR through a scalar field $\phi$. ``Galileon" means that the Lagrangian of the theory is invariant under the transformation $\phi\rightarrow\phi+b_{\mu}x^{\mu}+c$, and ``cubic" refers to the highest order of three in terms of $\phi$. Previous works have studied the properties of cubic and higher-order Galileon theories \citep{PhysRevLett.105.111301, 2011PhRvD..83d3515D, 2010PhRvD..82l4054N, Barreira_2013_056, Barreira_2013, 2013PhRvD..87j3511B, 2017JCAP...10..020R}, including linear perturbations and non-linear structure formations. Most of them consider a ``kinetic essence'' case, i.e. ignoring the first term, a potential term of the Lagrangian of the Galileon theory. In other words, in these studies, the acceleration of the universe is driven by the kinetic term. However, it is difficult to induce a stable late-time acceleration for CG without the potential term \citep{PhysRevD.102.043510, 2010PhRvD..82b4011G}. Meanwhile, the constraint on the gravitational wave propagation speed from the events of binary neutron star merger \citep{2017PhRvL.119p1101A,2017ApJ...848L..12A,2017ApJ...848L..13A} has ruled out a large class of Horndeski models \citep{2017PhRvL.119y1304E,2020PhRvD.102b3523Z}, including Galileon theories with terms whose order of the Galileon field $\phi$ is higher than three. Therefore, investigating a CG model with a potential term is more realistic for comparison with observations.

There are many difficulties in distinguishing MG theories from GR. One of them is the ``screening mechanism'', which suppresses the deviations from GR in high-density or small-scale regions due to the great success of GR in those environments. There are mainly two types of screening mechanisms. The first is called the ``chameleon mechanism'' \citep{2004PhRvD..69d4026K, 2004PhRvL..93q1104K}, which suppresses the observational effects of MGs in high-density regions where the Newtonian potential is deep enough. The other is the ``Vainshtein mechanism'' \citep{1972PhLB...39..393V, 2013CQGra..30r4001B}, which appears in CG theory and introduces high-order derivative interactions to suppress modification terms on small scales. These mechanisms allow MGs to pass local tests in the solar system, but prevent us from distinguishing them from GRs in these regions.

In contrast, cosmic voids, the underdense regions in large-scale structures, provide a special environment to explore the differences between GR and MG theories. Due to the properties of ``underdensity" and large scales, voids are less influenced by both the chameleon and Vainshtein mechanisms, making them an ideal probe for testing MG theories. For instance, \citet{10.1093/mnras/stv777}, \citet{2019A&A...632A..52P} and \citet{2019MNRAS.490.4907D} have examined several void properties such as void abundance, void density profile, and void lensing profiles in MG theories. \citet{2018MNRAS.476.3195C} and \citet{2019MNRAS.484.1149P} have investigated these properties of voids on chameleon-screened and Vainshtein-screened gravity theories, respectively, but focus on the differences between various void finders. Several works have used a single property of void to test a special MG theory. For example, \citet{Barreira_2015_08} and \citet{PhysRevD.98.023511} focused on void lensing in CG theory; \citet{2017JCAP...02..031B} also focused on void lensing but for the normal branch of
Dvali-Gabadadze-Porrati braneworld theory (nDGP), and \citet{2021PhRvD.104b3512W} used the void velocity profile to test $f(R)$ gravity. For a special case, \citet{2022MNRAS.515.5358C} studied the properties of galaxies in voids and observed some fingerprints of the $f(R)$ theory. Results from these researches support that void is a useful tool for studying and testing MG theories with screen mechanisms.

Weak lensing by cosmic voids is a powerful tool to test gravity. Observationally, weak lensing effects can be detected by measuring the distortion of the shape of background galaxies. Thus, it is convenient to compare lensing effects between theory and observation/simulation. From a theoretical perspective, lensing signals are sensitive to the lensing potential, which is directly influenced by modifications of gravity. This makes weak lensing a good probe of deviations from GR. In MGs, lensing signals can be influenced by both the dark matter distribution\footnote{{In this work we do not consider the baryon effect, therefore effects due to the change of the structure formation directly result in the differences of the dark matter field between two gravity models. Meanwhile, in this work, we focus on void lensing, where voids are chosen as the tracer of dark matter. Therefore, the change in the dark matter field causes different void density profiles. In this paper there are different expressions to describe this effect: ``Change of the structure formation'', ``Change of the dark matter field'' and ``Change of the void profile'', and we point out that these are all equivalent.}} and the deflection angle, which has been observed in the literature for a long time \citep{Barreira_2015_08, 2017JCAP...02..031B, PhysRevD.98.023511, 2018IJMPD..2748007C}, and is always considered in theoretical calculations. Nevertheless, when measuring lensing signals from simulations, the case becomes more complex. In GR, the lensing convergence can directly represent the distribution of matter density, as there is a linear relation between the second derivative of Newtonian potential and matter density. This coincidence brings the viewpoint that the lensing tangential shear (obtained by differentiating the convergence) represents the excess surface mass density. By projecting the 3D dark matter density from N-body simulation, one can obtain the tangential shear signals. However, in most MG theories, the relation between Newtonian potential and matter density is no longer linear, breaking the coincidence. Therefore, the lensing signal cannot be obtained from the projected differential mass density as it does not account for the effect due to the modification of the deflection angle.

In this work, we employ ray-tracing methods to analyze the lensing signals around voids in the context of CG theory. With the ray-tracing simulation results, the (mock) source galaxy shear catalog can be constructed. Combining this with the lens void catalog, we can measure the lensing signals around the voids. Our methodology has several advantages compared with the previous works, including the consideration of modifications to both the structure formation and the deflection angle (compared with directly projecting the 3D density profile), and the full contributions from the matter between the source plane and the observer (compared with cutting part of the region near the lens objects). In this sense, our results are more reliable, motivating the first scope of this work that is to investigate if those two approximate strategies are valid for CG theory. One another scope of this work is to test the possibility (and if possible, capability) of void lensing in distinguishing CG and GR. 

The structure of the paper is as follows: in Sec.~\ref{sec:theory} we briefly introduce the theory investigated in this work, which contains the gravity model and basic equations in the lensing calculation. In Sec.~\ref{sec:simu} we introduce our simulation information, in which the void properties used in the following section are also described and discussed. In Sec.~\ref{sec:result} we present our main results, including theoretical calculations and simulation results. The capability of void lensing to distinguish MG models is also discussed in this section. And Sec.~\ref{sec:sum&conclu} is the summary and conclusions of this paper.

\section{Theory} \label{sec:theory}
In this section, we will outline the theoretical framework employed in this work. First, we present the basic theory of the cubic Galileon gravity model, then we generally describe the calculation procedure of the lensing signal. At the end of this section, we discuss the subtleties of the lensing statistics, $\Delta\Sigma$, illustrating that it may not be the physical surface mass density in MG theories. The following discussion is based on the perturbed Friedmann-Robertson-Walker (FRW) metric with the Newtonian gauge
\begin{equation}
    \mathrm{d}s^2= -(1+2\Psi)\mathrm{d}t^2 + a^2(t)(1-2\Phi)\gamma_{ij}\mathrm{d}x^i\mathrm{d}x^j, \label{metric}
\end{equation}
where $a=1/(1+z)$ is the scale factor with the redshift $z$. The background FRW metric satisfies $\Phi=\Psi=0$.
\subsection{Gravity model}
Galileon theory modifies GR by introducing a scalar field, $\phi$, with its unusual high-order derivative self-interaction \citep{PhysRevD.79.064036}. The action of Galileon gravity is
\begin{align}
S=& \int d^{4} x \sqrt{-g}\left[\frac{M_{\mathrm{pl}}^{2}}{2} R-\frac{1}{2}(\nabla \phi)^{2}(1+\beta \square \phi)-V(\phi)\right]\nonumber \\
&+\mathcal{S}_{m}, \label{action}
\end{align}
where $M^3=M_{\text{Pl}}H_0^2$, $M_{\text{Pl}}=1/\sqrt{8{\pi}G}$ is the reduced Planck mass. The variation of the action w.r.t. metric tensor $g_{\mu\nu}$ in flat FRW metric leads to two background Friedmann equations:
\begin{align}
3 M_{\mathrm{pl}}^{2} H^{2}=\bar{\rho}_{\mathrm{m}}+\frac{\dot{\phi}^{2}}{2}(1-6 \beta H \dot{\phi})+V(\phi) \\
M_{\mathrm{pl}}^{2}\left(2 \dot{H}+3 H^{2}\right)=-\frac{\dot{\phi}^{2}}{2}(1+2 \beta \ddot{\phi})+V(\phi). \label{Frdm}
\end{align}

Meanwhile, the equation of motion of galileon field $\phi$ can be derived by varying the action w.r.t. $\phi$. We split the full galileon field $\phi$ into homogeneous part $\bar{\phi}$ and the perturbed part $\varphi$. In flat FRW metric the equation of motion of $\bar{\phi}$ reads:

\begin{equation}
\ddot{\bar{\phi}}+3 H \dot{\bar{\phi}}-3 \beta \dot{\bar{\phi}}\left(3 H^{2} \dot{\bar{\phi}}+\dot{H} \dot{\bar{\phi}}+2 H \ddot{\bar{\phi}}\right)+V_{\phi}=0, \label{eom}
\end{equation}
where overdots represent derivations with physical time and $V_{\phi}$ represents the derivation w.r.t. $\bar{\phi}$. Using a set of dimensionless quantities to replace the variables in Eqs.~\ref{Frdm} and \ref{eom}, these equations can be solved numerically \citep{PhysRevD.102.043510,HOSSAIN2012140}. The initial value of one of the dimensionless quantities, $\epsilon=-6{\beta}H^2\phi^{\prime}$, where a prime represents the derivative w.r.t. the number of e-foldings $N\equiv\ln(a)$, can be chosen as a free parameter. However, in this work we fix this free parameter $\epsilon_i$ and regard the CG theory as a model without any free parameters, leaving constraining the parameter to future work. We choose $\epsilon_i=0$ because in this case, CG and GR universes present the most distinct properties.

Furthermore, it is necessary to perform perturbation calculations, since the lensing signal is sensitive to the perturbation quantities in the metric tensors. Varying the action Eq.~\ref{action} w.r.t. the metric and the Galileon field, respectively, and in the perturbed FRW metric Eq.~\ref{metric} we can obtain the perturbed Einstein equations as well as the equation of the motion of the perturbed Galileon field $\delta\varphi$ up to the first order. These tedious equations can be found in \citet{PhysRevD.102.043510} and are not shown here. Finally, on the strength of the evolution of $\Phi$ and the perturbed metric potentials, we are able to calculate the lensing signal.

\subsection{Weak lensing signal} \label{lencal}
The weak lensing effect originates from light bending by foreground objects or structures. This effect can cause the distortion of the shapes of background galaxies, which can be quantified by the tangential shear $\gamma_t$. It is related to lensing convergence, $\kappa$ by: 
\begin{equation}
    \gamma_{\text{t}}=\bar{\kappa}-\kappa, \label{gt_kappa}
\end{equation}
where $\bar{\kappa}$ represents the mean convergence in a given radius, $\bar{\kappa}\left(<R\right)=\int_{0}^{R}{r{\kappa}\left(r\right)\mathrm{d}r}$.

Theoretically, $\kappa$ is half of the second derivation of the lensing potential \citep{2006glsw.conf....1S}, which is the integral of the sum of the metric potentials, i.e. lensing potential:
\begin{align}
    \kappa&=\frac{1}{2}\nabla^2\psi_{\text{lens}}=\frac{1}{4{\pi}G{\Sigma_{\text{crit}}}}\int{\frac{1}{2}\nabla^2\left(\Phi+\Psi\right)}\mathrm{d}\ell, \label{kappa}
\end{align}
where $\Sigma_{\text{crit}}$ is the geometry factor dependent on the lens system. $\ell$ represents the comoving distance along the line of sight. For GR, $\Phi=\Psi$, and by using the Poisson equation, $\nabla^2\Phi=4{\pi}Ga^2\rho_m\delta$, Eq.~\ref{kappa} can be expressed as:
\begin{align}
    \kappa&=\frac{1}{4{\pi}G\Sigma_{\text{crit}}}\int{4{\pi}Ga^2\rho_m\delta\mathrm{d}\ell} \nonumber\\
    &=\frac{\bar{\rho}_{m0}}{\Sigma_{\text{crit}}}\int{\frac{\delta}{a}\mathrm{d}\ell}=\frac{\bar{\rho}_{m0}}{\Sigma_{\text{crit,c}}}\int{\delta\mathrm{d}\ell},
\end{align}
where for the second line we use ${\rho}_m=\bar{\rho}_{m0}a^{-3}$. $\Sigma_{\text{crit}}$ and $\Sigma_{\text{crit,c}}$ represent the geometry factor expressed in physical and comoving coordinates, respectively. Note that $\Sigma_{\text{crit}}$ has the meaning of surface mass, thus, analogous to critical density $\bar{\rho}_m$, with the expansion of the universe, $\Sigma_{\text{crit}}$ should be inversely proportional to $a^{-2}$. In other word, the relation between the expression of $\Sigma_{\text{crit}}$ in physical coordinates and in comoving coordinates is $\Sigma_{\text{crit}}=\Sigma_{\text{crit,c}}{\times}a^{-2}$ \citep{10.1093/mnras/sty1624}.

For CG, there is no anisotropic stress, i.e. $\Phi=\Psi$, however, $\Phi_{CG}\neq\Phi_{GR}$. In this case, we rewrite the Poisson equation as $-k^2\Phi=4{\pi}a^2G_{\text{eff}}(k,a)\delta\rho_k$, and first apply a Fourier transform on void density profile to obtain $\rho_k$, then calculate $-k^2\Phi$ and finally apply an inverse Fourier transformation to get $\nabla^2\Phi$.

With convergence $\kappa$, the excess surface mass density (ESD) can be given as:
\begin{equation}
    \Delta{\Sigma}=\Sigma_{\text{crit,c}}\left(\bar{\kappa}-\kappa\right). \label{esd_kappa}
\end{equation}
From Eq.~\ref{esd_kappa} we can see that the calculation of $\Sigma_{\text{crit,c}}$ can be avoided. The only ingredient needed in the calculation is the density profile of the foreground objects, i.e. the void density profile in this work. Instead of using a fitting function, we directly use the profile from the N-body simulation, since the void profiles in the two gravity models are very similar \citep{2018MNRAS.476.3195C}. Details of the void profile can be found in section \ref{sec:simu}.

\subsection{Discussions on ESD}\label{choose}
In the end of this section, we aim to provide a discussion on ESD, $\Delta\Sigma$, a commonly used statistics for studying weak lensing effects in previous literature \citep{10.1093/mnras/stu456, 10.1093/mnras/stv777, Barreira_2015_08, 10.1093/mnras/stv2215}. It should be noted that as a lensing statistic, $\Delta\Sigma$ may not have the physical meaning of excess surface mass density. The original definition of the surface mass density is simply the integral of the volume density along one dimension, $\Sigma=\int{\rho\mathrm{d}\ell}$. In GR, the integral part of Eq.~\ref{kappa} is equivalent to the surface mass density by coincidence, since the second deviation of the Newtonian potential is proportional to the volume density. However, in MGs one needs to subtract the effects from the modified theory when calculating the (real) surface mass density from the lensing statistics $\Delta\Sigma$. For instance, in a certain MG theory where the Poisson equation reads $\nabla^2\Phi_{\text{MG}}=4{\pi}G{\rho}+\mathcal{M}$, the surface mass density should be expressed as $\Sigma=(1/4{\pi}G)\int{(\nabla^2\Phi_{\text{MG}}-\mathcal{M})\mathrm{d}\ell}$ 
rather than $\Sigma=(1/4{\pi}G)\int{\nabla^2\Phi_{\text{MG}}\mathrm{d}\ell}$, in order to be consistent with its original definition. Therefore, directly projecting the 3D dark matter density may not induce the correct lensing signal. To this end, we employ the ray-tracing algorithm to simulate the observed convergence map and shear map in the scenario of CG, and measure the lensing signal directly from the simulation results. Certainly, one salient advantage of this approach is its independence of the gravity framework in which one works, implying its flexible usage in any MG theory.

\section{Simulation} \label{sec:simu}
In this section, we describe the ingredients necessary for the theoretical calculation and measurement of the lensing signal. Specifically, we require the density profile of the foreground void as input, as well as the background shear and foreground void catalogs for the measurements. We begin with introducing our N-body simulation and subsequently construct the void catalog. Next, we measure the void density profiles in both gravity models. Finally, we provide a detailed description of our ray-tracing method, which enables us to obtain the observed convergence and shear maps from the simulated void and matter density fields.

\subsection{N-body simulation}\label{nbody}
We use the \textsc{me-gadget} code \citep{PhysRevD.102.043510, 2018PhRvD..98j3530Z} to generate the simulation of $512^3$ particles in a periodic box whose size is $L=400~\rm Mpc/h$ for both GR and CG simulations, with a softening length of $25~\rm kpc/h$. The initial condition is generated with 2LPTic \citep{2006MNRAS.373..369C} while the pre-initial condition file is generated by CCVT \citep{2018MNRAS.481.3750L}. For the GR case, we assume a Flat $\Lambda$CDM model with a cosmological parameter set of $\left\{\Omega_{m0},\Omega_{b0}, h, \sigma_8, n_s\right\}=\left\{0.3156, 0.0491, 0.6727, 0.831, 1.0 \right\}$. For the CG case, we choose $\epsilon_i=0$ which is the most different case from GR. We store 11 snapshots from $z=0$ to $z=1$ with a redshift interval of ${\Delta}z=0.1$, poised to generate the backward light cones in the next step. In order to reduce the cosmic variance in the following measurements, for each gravity model we keep the cosmological parameters the same but change the random seeds to generate 10 realizations and average the results obtained from those simulations.

\begin{figure*}
    \centering
    \includegraphics[scale=0.5]{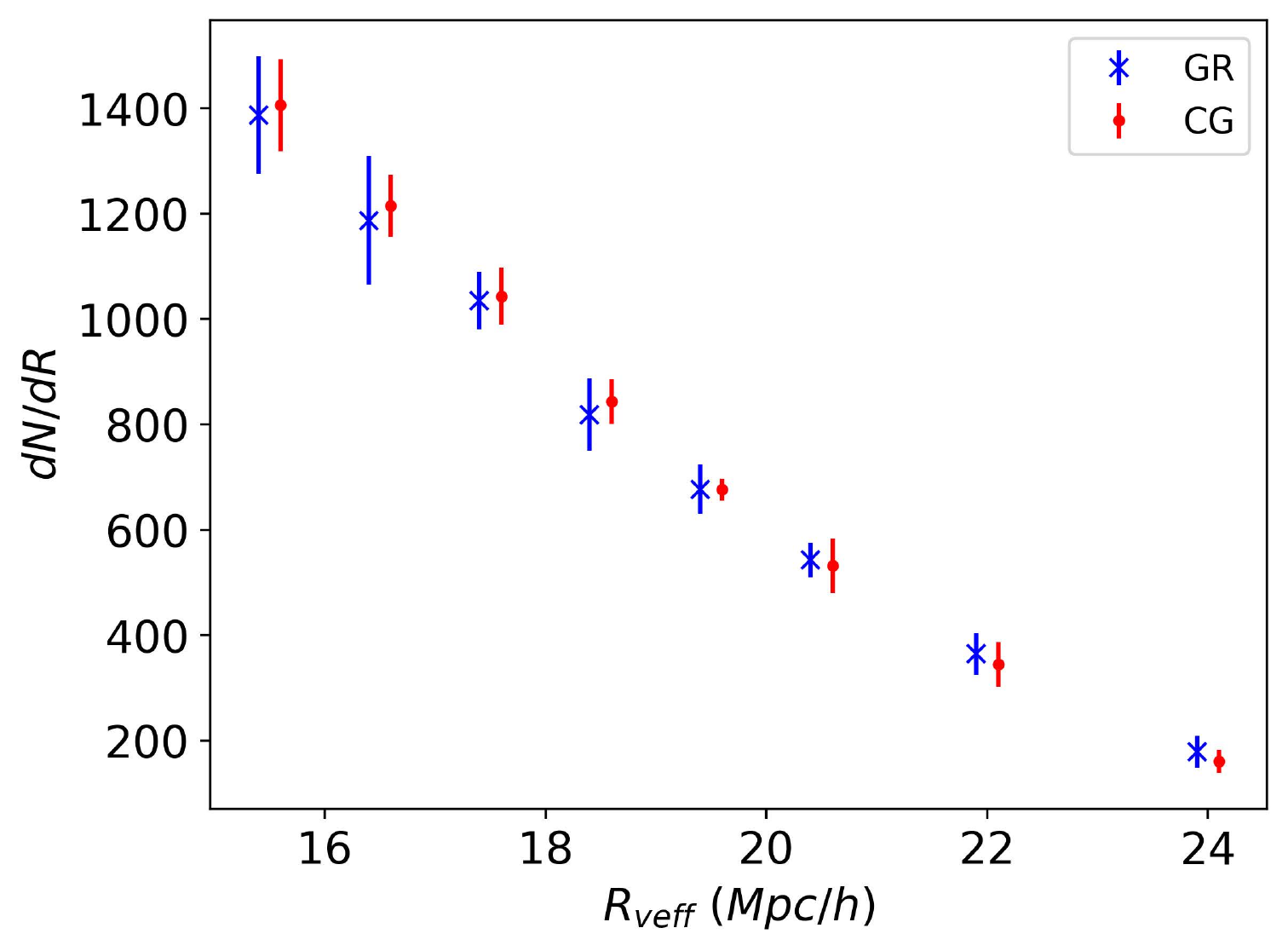}
    \caption{The void size function in two gravity models. The red dots represent GR voids and the blue cross points represent CG voids. The error bar represents sample variance. It can be seen that voids in the two simulations have a similar radius distribution.}
    \label{Rbins}
\end{figure*}

\subsection{Void catalog and density profile}\label{vdens}
In this work, we rely on Delaunay TrIangulation Void findEr (\textsc{dive}, \citet{2016MNRAS.459.2670Z}) to identify voids from galaxy samples. First, we generate the halo catalogs from N-body simulation by using \textsc{ahf} halo finder \citep{2011ascl.soft02009K}. We then apply the SubHalo Abundance Matching (SHAM) method on the simulations to generate realistic mock galaxies that reproduce the observed clustering, thus providing a good sample to find voids as the lenses in the foreground. SHAM assigns galaxies to halos/subhalos based on the assumption of a monotonic galaxy luminosity-halo mass relation \citep{2004ApJ...609...35K, 2010ApJ...717..379B}. We generate mock galaxies with an effective redshift $z_{\text{eff}}=0.3$, as the redshift of lens mocks is sufficiently distant from the source to produce a significant lensing signal. The SHAM algorithm used in this work is described in \citet{2022MNRAS.516...57Y}. We fit the two-point correlation function of the SHAM galaxy catalog to that of the BOSS LOWZ samples at $0.2<z<0.33$. As we have 10 realizations of the N-body simulation, we apply the same parameter set to all realizations and average their output 2PCFs to obtain the best-fit SHAM parameters. The parameter set with the maximum likelihood is then implemented on all the realizations to generate the corresponding galaxy mocks with the same cosmology. Finally, we run {\footnotesize DIVE} on the SHAM galaxy samples to generate void catalogs. We divide the voids into 8 size bins based on their radius, ranging from $R_v=15~\text{Mpc/h}$ to $R_v=25~\text{Mpc/h}$. The interval of the first 6 bins is $1~\text{Mpc/h}$, while the last 2 bins have intervals of $2~\text{Mpc/h}$ since the number of voids with radius $R_v>21~\text{Mpc/h}$ is too small. Voids within the same size bin are assigned the same effective radius, defined by the median of the edge of the bin. For example, voids within the size bin of $15~\text{Mpc/h}<R_v<16~\text{Mpc/h}$ are assigned the same effective radius $R_{\text{veff}}=15.5~\text{Mpc/h}$. Fig.~\ref{Rbins} displays the void number distribution w.r.t. radius (${\mathrm{d}N/\mathrm{d}R}$) in simulations under GR and CG gravity, respectively, illustrating that the distributions of void radii are quite similar in both gravity models. These voids are regarded as lenses in the measurements of ESD in the subsequent section.

Void profiles are measured by cross-correlating the position of the center of voids and the dark matter particles, which can be estimated through pair counts, i.e. histograms binned by pair separations \citep{2022MNRAS.511.5492Z}. We use a simple estimator:
\begin{equation}
    \widehat{\xi}_{vm}=\frac{DD}{RR}-1,
\end{equation}
where $D$ and $R$ represent the data and random catalog, respectively. In this work \textit{data} refer to the catalog in simulations. To increase the number of voids for measuring the void lensing signal, we stack those voids with different effective radii. Practically, we rescale the coordinates of the tracers using the void effective radii and measure the pair counts as a function of rescaled distance. As voids with different effective radius should be assigned by different rescaled values, we measure the pair counts separately for the 8 size bins and obtain 8 sets of results. These pair counts are then added up to construct total pair counts. Finally, we estimate the void-dark matter cross-correlation function using these total pair counts. The Fast Correlation Function Calculator (\textsc{fcfc}, \citet{2021MNRAS.503.1149Z}) is used to measure the pair counts in individual size bins.

\begin{figure*}[t]
\centering  
\subfigure{
\includegraphics[scale=0.5]{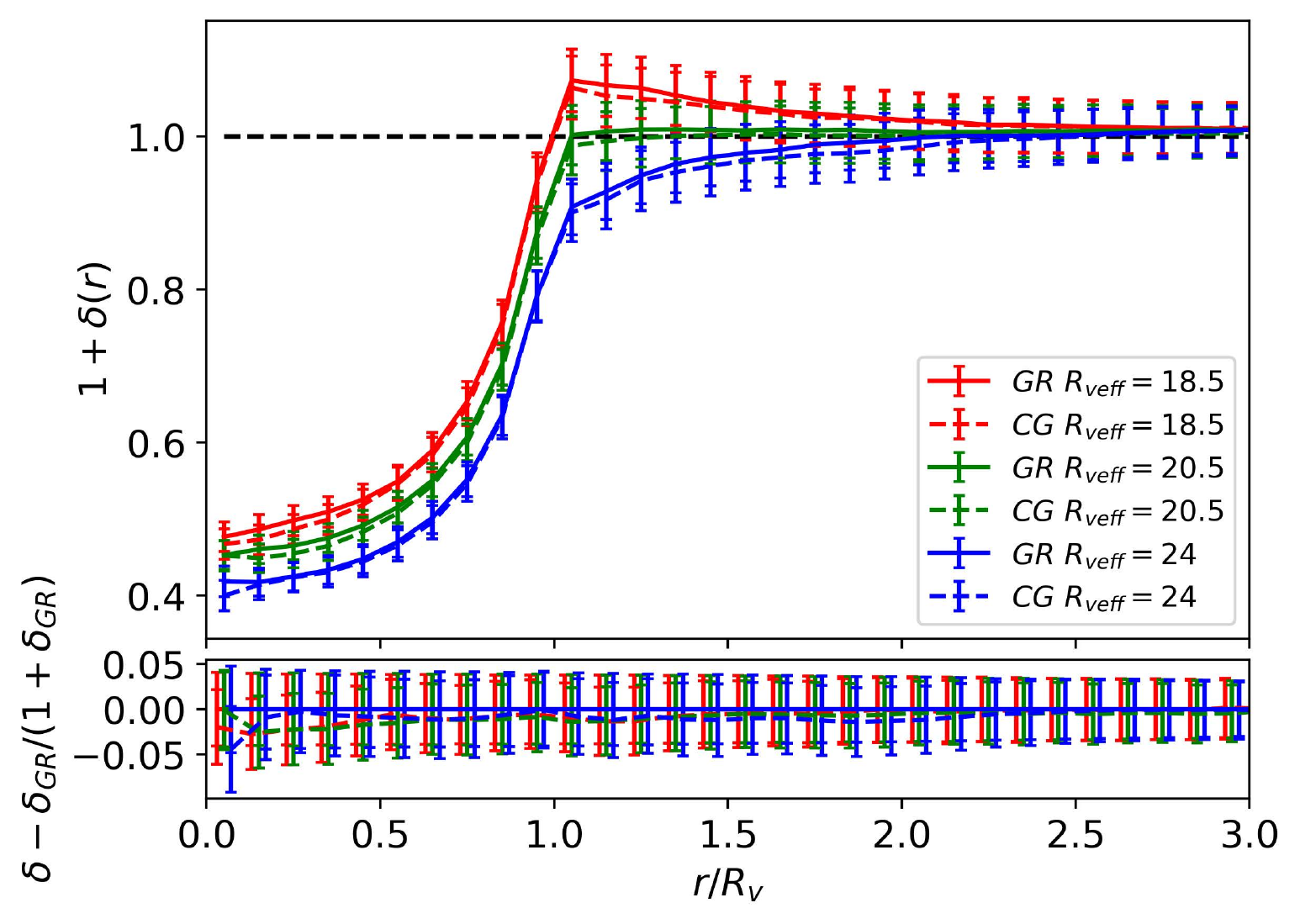}}\subfigure{
\includegraphics[scale=0.5]{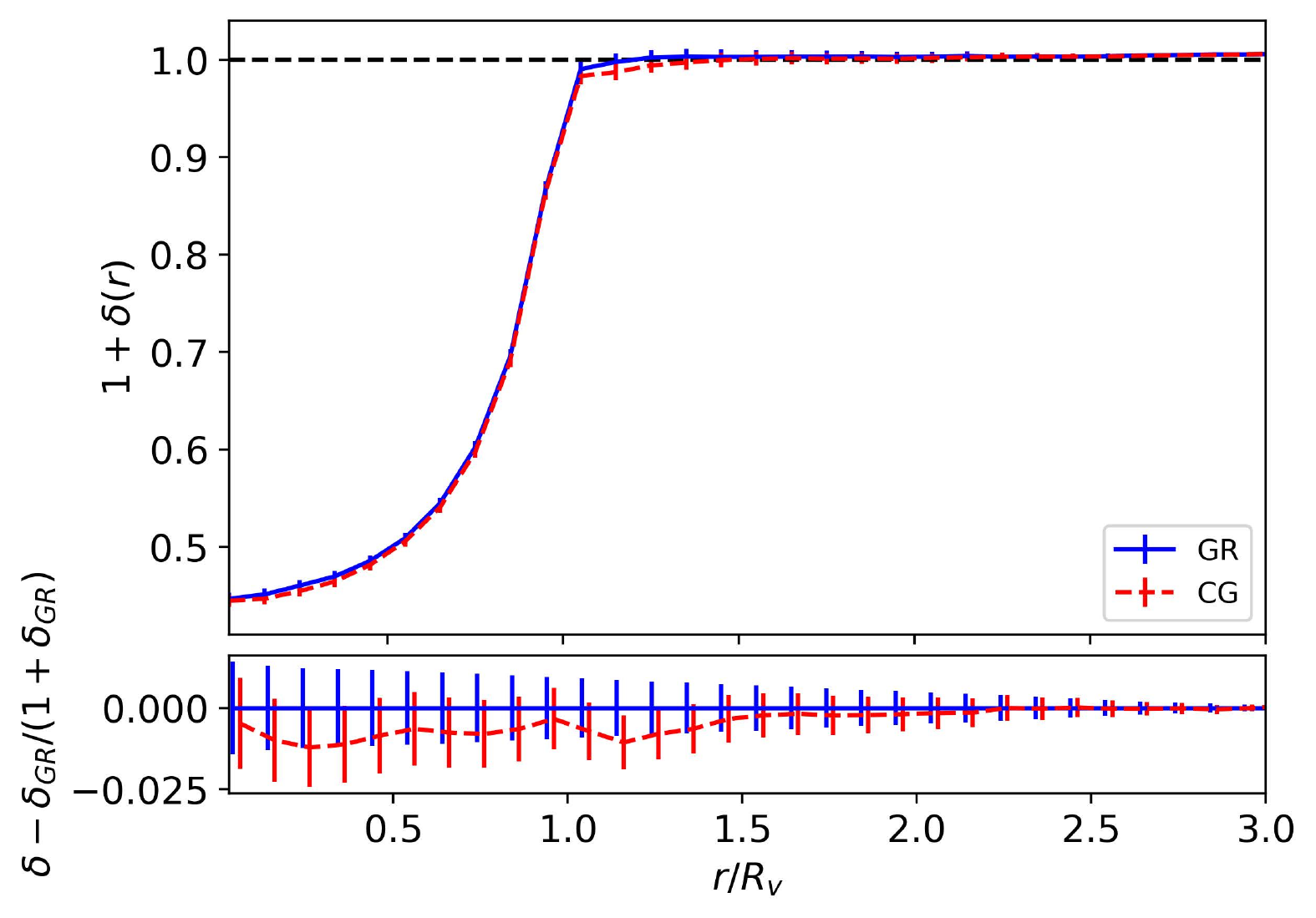}}
\caption{Void profiles as a function of the re-scaled distance to the void center, $r/R_v$. {Note that data points in the bottom panel are shifted to make the figure clearer.} \textit{Left}: Void density profiles for 3 void size bins: $18<R_v<19$, $20<R_v<21$, and $23<R_v<25$. The solid lines represent the GR void profiles, and the dashed lines represent the CG void profiles. One kind of color represents one void-size bin. {The error bars are estimated by jackknife technique.} It is obvious that voids of different sizes have different properties in density profile{, However, the differences between gravity models are not statistically significant.} \textit{Right}: Stacked void profile in two gravity models. The red dashed line represents the stacked CG void profile, and the blue solid line represents the stacked GR void profile. The errorbar is propagated from the jackknife error in a single simulation during the averaging of the void profile. One can see that the deviation between the two models is much smaller than in voids in a narrow size bin{, and is also not statistically significant.}}
\label{void_prof}
\end{figure*}

We average over the void profiles measured from 9 realizations for GR and CG cases, shown in Fig.~\ref{void_prof}. Only three radius size bins on the left panel are presented for the sake of clarity. The right panel shows the stacked void profile, where only the voids with a radius greater than $18~\text{\text{Mpc/h}}$ are stacked as the lensing signals around these voids are stronger. The differences in void profiles (relative to GR) are shown in the lower panels. From the left panel, one can see that the CG voids in all three size bins are slightly emptier than the GR voids, inferring the same characteristics in the lensing profile. It is also clear from the left panel of Fig.~\ref{void_prof} that voids with different sizes present different shapes of profiles. The larger the void, the emptier it is. Meanwhile, smaller voids tend to have a higher well near the radius of $r=R_{\text{veff}}$. The profile of the stacked voids exhibits a smoother shape, both for the GR and CG voids, as can be seen in the right panel of Fig.~\ref{void_prof}. Additionally, we observe a local valley at $r{\sim}1.1~R_{\text{veff}}$. {However, combined with the length of the error bar, these differences are not significant statistically. These void profiles will be used in the theoretical calculations of ESD in the next section.} In the last part of the paper, we would use $R_v$ to represent the effective radius $R_{\text{veff}}$.

\subsection{Ray-tracing algorithm}\label{ray}
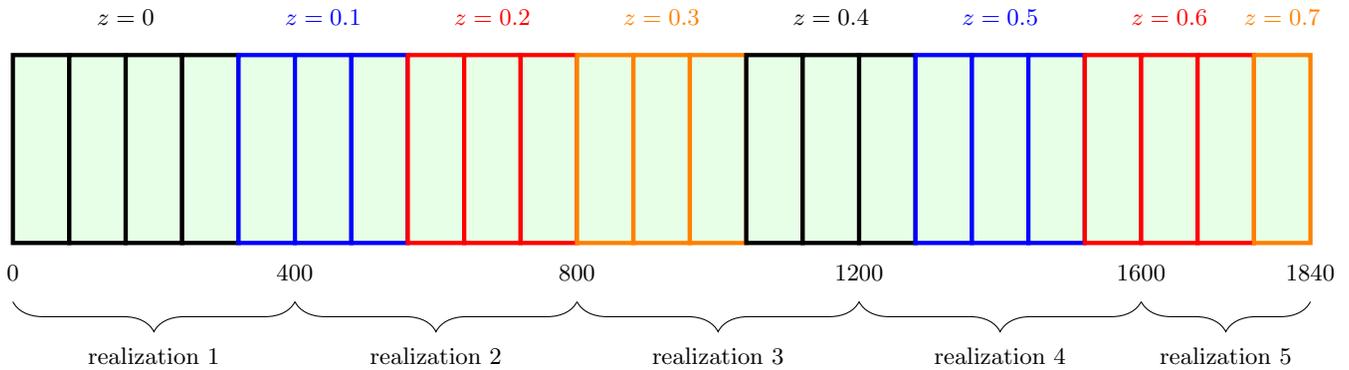
\begin{figure*}
    \centering
    \begin{tikzpicture}
        \draw (0.75*2.5, 0) node {realization 1};
        \draw (0.75*7.5, 0) node {realization 2};
        \draw (0.75*12.5, 0) node {realization 3};
        \draw (0.75*17.5, 0) node {realization 4};
        \draw (0.75*21.5, 0) node {realization 5};
        \foreach \i in {0,...,3}{
            \filldraw[shift={(0.75*\i, 0)}][fill=green!10!white,draw=black][ultra thick] (0,1.5) rectangle (0.75,4);
        }
        \foreach \i in {4,...,6}{
            \filldraw[shift={(0.75*\i, 0)}][fill=green!10!white,draw=blue][ultra thick] (0,1.5) rectangle (0.75,4);
        }
        \foreach \i in {7,...,9}{
            \filldraw[shift={(0.75*\i, 0)}][fill=green!10!white,draw=red][ultra thick] (0,1.5) rectangle (0.75,4);
        }
        \foreach \i in {10,...,12}{
            \filldraw[shift={(0.75*\i, 0)}][fill=green!10!white,draw=orange][ultra thick] (0,1.5) rectangle (0.75,4);
        }
        \foreach \i in {13,...,15}{
            \filldraw[shift={(0.75*\i, 0)}][fill=green!10!white,draw=black][ultra thick] (0,1.5) rectangle (0.75,4);
        }
        \foreach \i in {16,...,18}{
            \filldraw[shift={(0.75*\i, 0)}][fill=green!10!white,draw=blue][ultra thick] (0,1.5) rectangle (0.75,4);
        }
        \foreach \i in {19,...,21}{
            \filldraw[shift={(0.75*\i, 0)}][fill=green!10!white,draw=red][ultra thick] (0,1.5) rectangle (0.75,4);
        }
        \filldraw[shift={(0.75*22, 0)}][fill=green!10!white,draw=orange][ultra thick] (0,1.5) rectangle (0.75,4);
        \foreach \i/\j in {5/0, 10/5, 15/10, 20/15, 23/20}{
            \draw[decorate,decoration={brace,raise=8pt,amplitude=0.4cm},black] (0.75*\i,1) -- (0.75*\j,1);
        }
        \foreach \i/\j in {0/0,400/5, 800/10, 1200/15, 1600/20, 1840/23}{
            \draw (0.75*\j, 1.1) node {\i};
        }
        \draw[ultra thick, black] (0.75*2, 4.5) node {$z=0$};
        \draw[ultra thick, blue] (0.75*5.5, 4.5) node {$z=0.1$};
        \draw[ultra thick, red] (0.75*8.5, 4.5) node {$z=0.2$};
        \draw[ultra thick, orange] (0.75*11.5, 4.5) node {$z=0.3$};
        \draw[ultra thick, black] (0.75*14.5, 4.5) node {$z=0.4$};
        \draw[ultra thick, blue] (0.75*17.5, 4.5) node {$z=0.5$};
        \draw[ultra thick, red] (0.75*20.5, 4.5) node {$z=0.6$};
        \draw[ultra thick, orange] (0.75*22.5, 4.5) node {$z=0.7$};
    \end{tikzpicture}

    \caption{A schematic diagram of the lightcone. Each rectangle represents a slice. The colors of the slices represent the redshifts of snapshots used to fill the slice. The comoving distances for the edges of slices are in Mpc/h. We also mark the realizations used in constructing the lightcone.}
    \label{lgtcn}
\end{figure*}

We trace the light rays from $z_{\text{source}}=0.75$ to $z_{\text{obs}}=0$, corresponding to a comoving distance of $d\approx1840\ \text{Mpc}/h$, and the simulated area is about $12.45\times12.45\ \text{deg}^2$. Our ray-tracing algorithm is mainly based on \citet{AA.499.1.2009}, and we briefly outline the main steps here. The backward lightcone is divided into 23 redshift slices, each of whose width is of $\ell=80~\text{Mpc}/h$. Since the slice width is smaller than the boxsize, we take every 3 slices from one snapshot, except for the first 4 and the last one. To avoid encountering the same structures many times along the line of sight, we change the realization of the snapshot every five slices (i.e., a total width equal to the box size of our simulation). A schematic diagram of the lightcone is illustrated in Fig.~\ref{lgtcn}. We construct 9 different lightcones by taking turns using the 10 N-body simulations, then employ the ray-tracing method on these lightcones. Finally, we average the measurements from these ray-tracing results to reduce the cosmic variance.

\begin{figure*}
    \centering
    \includegraphics[scale=0.8]{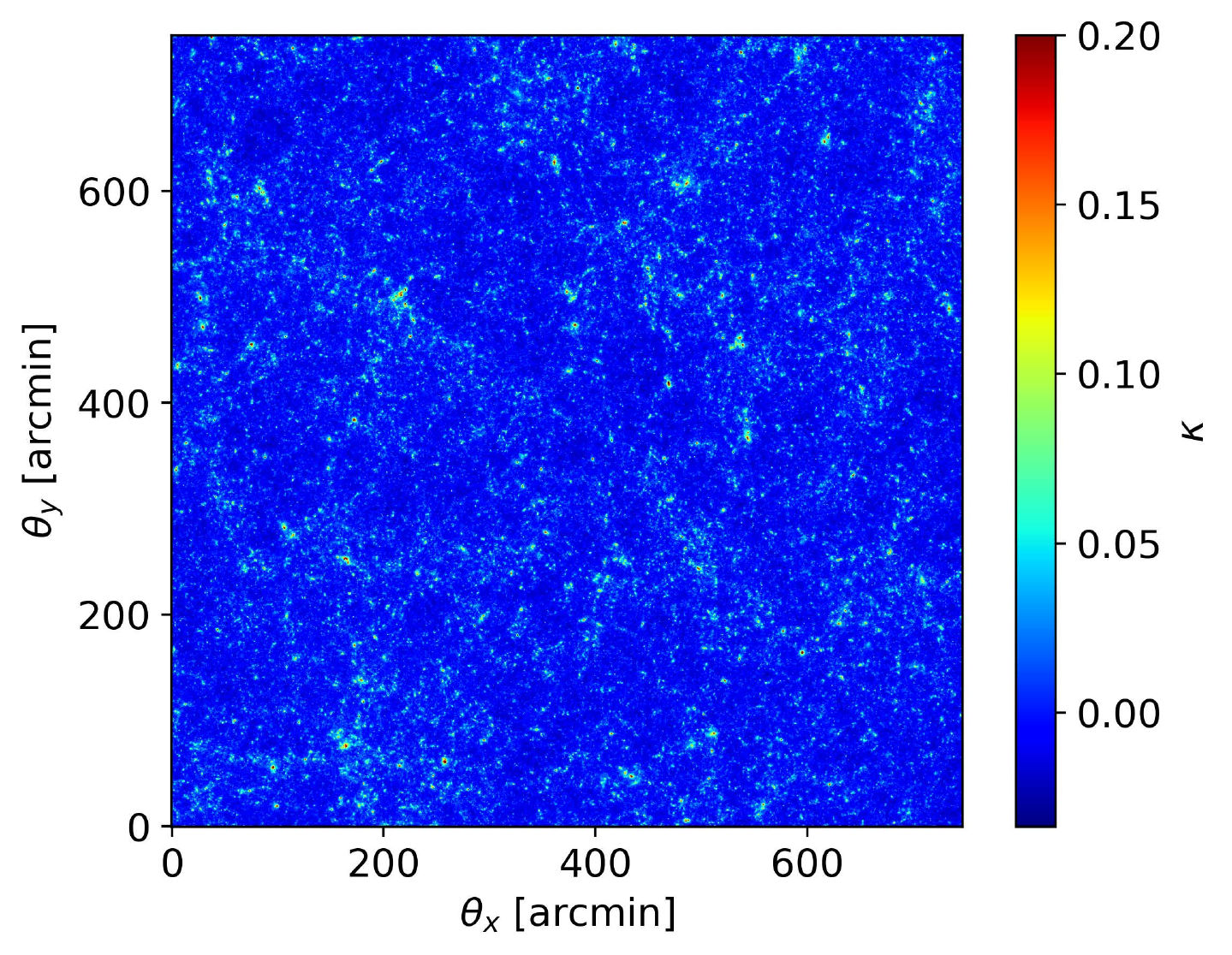}
    \caption{The simulated convergence map {for one realization based on GR gravity}. The color bar represents the amplitude of the convergence.}
    \label{kap_gam}
\end{figure*}

After constructing the lightcone, the lens planes are generated by projecting the particles in slices along the line of sight. The comoving distance of each lens plane is determined by the redshift of the corresponding slice. The matter density field is then constructed using the particle-mesh (PM) method, which is used to compute the gravitational potential. Finally, the deflection angle and the corresponding quantities, such as lensing convergence and shear, can be calculated straightforwardly. For CG, the Poisson equation can be rewritten as $-k^2\Phi=4{\pi}G_{\text{eff}}(k,z)\delta\rho_m$, and in principle we can replace the constant $G$ with $G_{\text{eff}}$ when calculating the deflection angle. However, since $G_{\text{eff}}$ is very close to the gravitational constant $G$, we choose a simpler way to test the influences of the modification to the Poisson equation. First, we determine the maximum value of $G_{\text{eff}}$ in the range of $k$ and $z$ of interest. Then, the original gravitational constant $G$ is directly replaced with this maximal value. We believe that, in this case, the deflection angle is influenced to the most extent. If the difference between the two ray-tracing results is indistinguishable, it is unnecessary to calculate the real deflection angle in the simulation.

\begin{figure*}[t]
\centering  
\subfigure{
\includegraphics[scale=0.5]{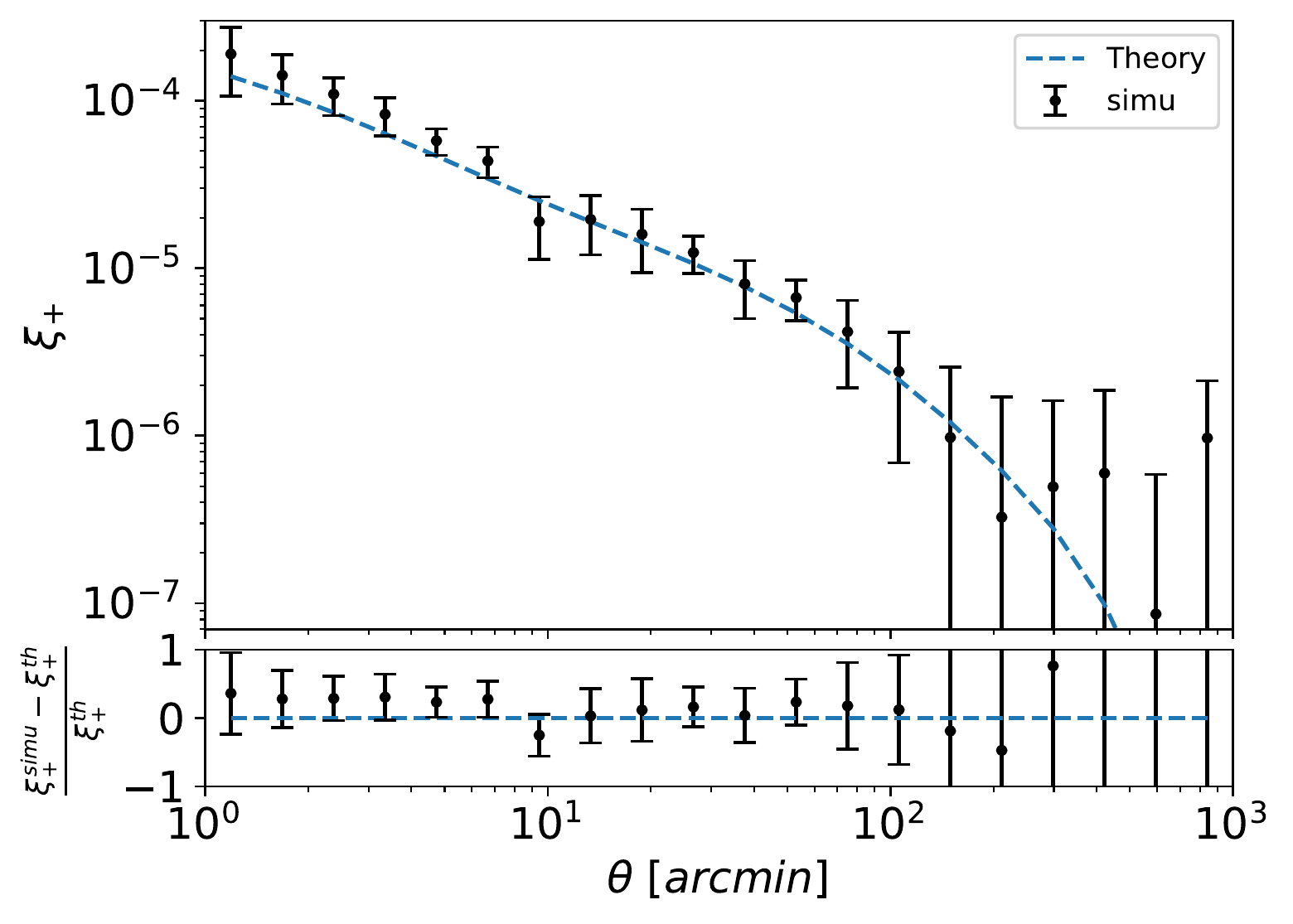}}\subfigure{
\includegraphics[scale=0.5]{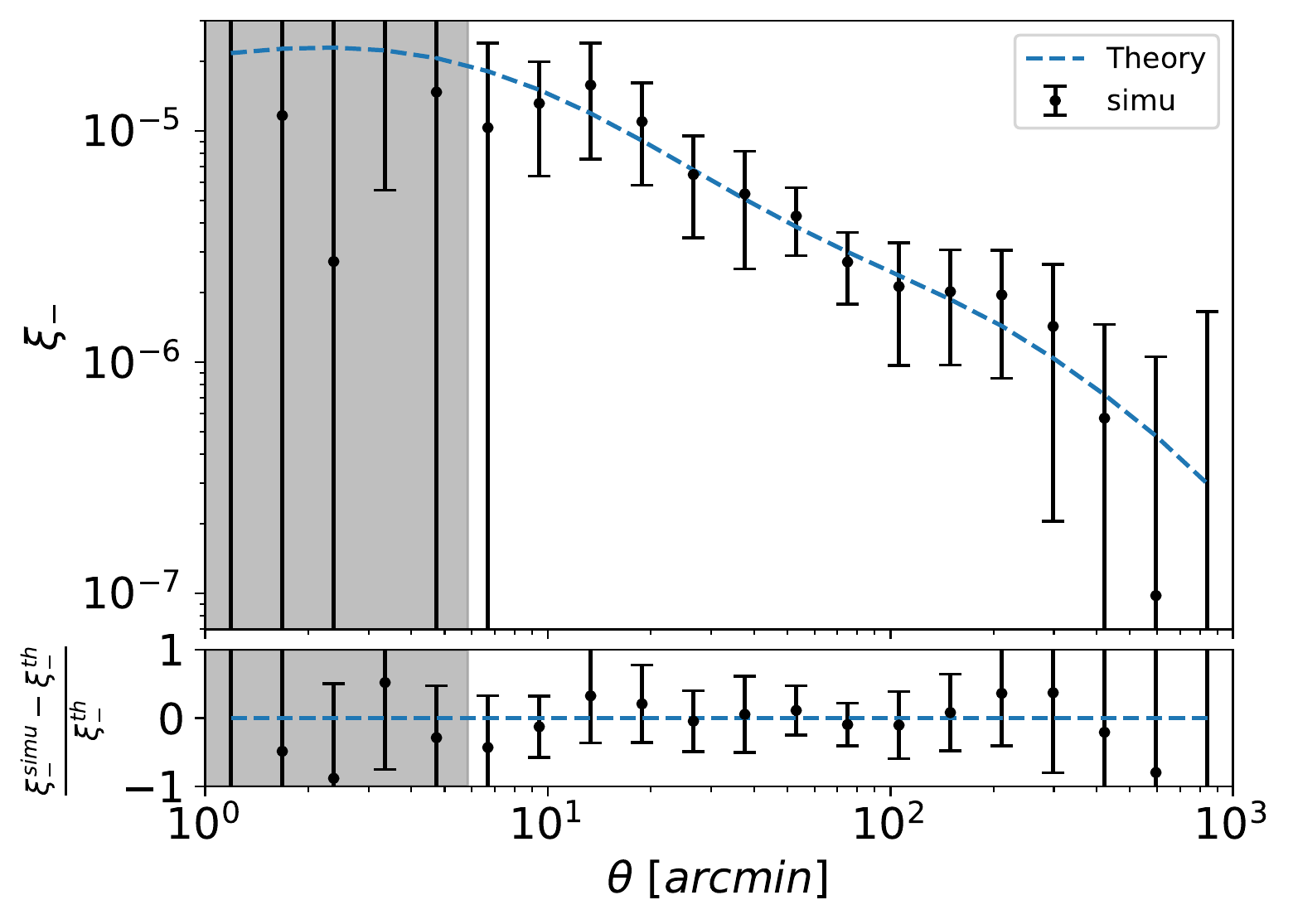}}
\caption{Cosmic shear statistics in GR simulation. The left panel shows $\xi_{+}$ and the right panel shows $\xi_{-}$. Relative differences between theory and simulation are presented in the bottom panels. The blue dashed line represents the theoretical prediction, while the black dots with error bars represent the measurements from the simulated shear map. The shaded region indicates the smoothed regions by PM method. One can see that simulation results agree with the theoretical predictions.}
\label{xires}
\end{figure*}

Fig.~\ref{kap_gam} presents the convergence field obtained from one single realization, in which the color bar represents the convergence (surface mass density). To validate our simulation results, we calculate the shear-shear correlation functions, $\xi_{+}$, and $\xi_{-}$ in the $\Lambda$CDM simulation and compare with the theoretical predictions. We randomly sample 558,282 galaxies (corresponding to the galaxy number density of $1~\text{arcmin}^{-2}$) and place them down on the surface at a redshift of $z=0.75$, resulting in a uniform distribution of their projected distances. Note that since this is only a test, the population of background galaxies is not expected to be realistic. However, in order to obtain a reliable void catalog we need to construct a more realistic galaxy catalog since the void finder is sensitive to the foreground galaxy catalog. This is the reason we use SHAM to populate galaxies in the previous section. We also consider the noisy shear components in the measurement by including a random ellipticity drawn from a Gaussian distribution with the standard deviation of $\sigma_e=0.3$ (see Sec.~\ref{vlsignal}). For the theoretical predictions, We use \textsc{nicera} \citep{2009A&A...497..677K} to calculate the theoretical cosmic shear. The non-linear matter power spectrum model is from \citet{2012ApJ...761..152T} and the transfer function is from \citet{1998ApJ...496..605E} with baryon acoustic oscillation (BAO) wiggles. The results are exhibited in Fig.~\ref{xires}, where the measurements are obtained by averaging the results from 9 lightcones. The shaded region marked in $\xi_{-}$ indicates the angular scale corresponding to the length of the mesh grid ($\sim6~\text{arcmin}$) chosen in the PM algorithm, where the measurements should not be reliable. In general, the results of our ray-tracing simulation are reliable. It should be noted that we use a different source of the nonlinear power spectrum, but we do the consistency test and find that the differences between the power spectra obtained from theory and simulation are smaller than 5\%, which is smaller than the cosmic variance. Considering the large error bar in Fig.~\ref{xires}, this discrepancy due to the usage of a different source of power spectrum does not influence the conclusion of this check.

\section{Results} \label{sec:result}
In this section, we present our main results both in theory and simulation. For theoretical calculations, we used the void profiles measured in Sec.~\ref{vdens}. The change of the dark matter field and the deflection angle are considered separately during the calculation. For simulation results, we cross-correlate the foreground void position field with the background shear field to obtain the lensing signal. Finally, we analyze the capability of the void lensing effect in distinguishing gravity models.

\subsection{Theoretical lensing signal}\label{prof}
\begin{figure*}[t]
\centering  
\subfigure{
\includegraphics[scale=0.5]{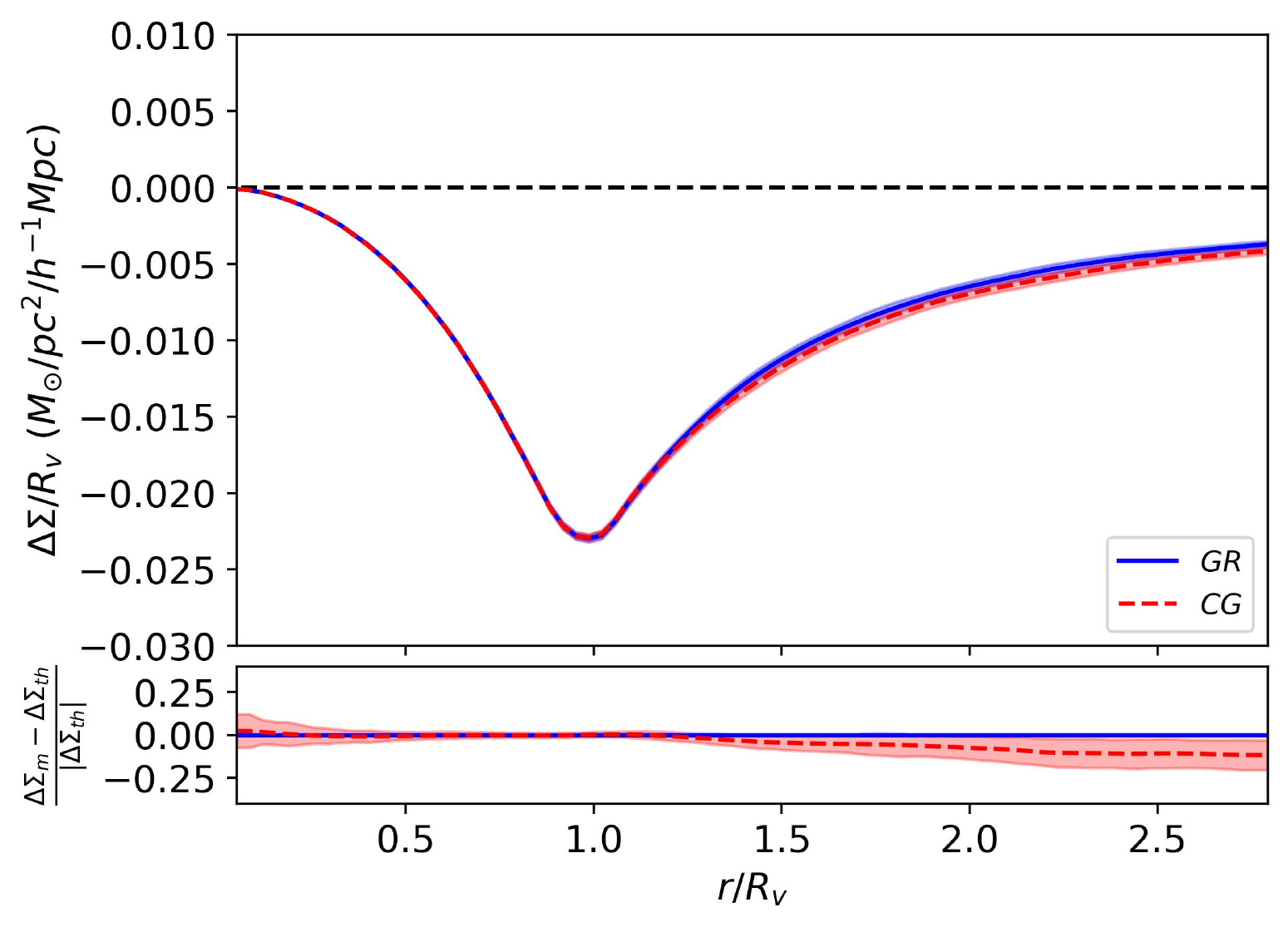}}\subfigure{
\includegraphics[scale=0.5]{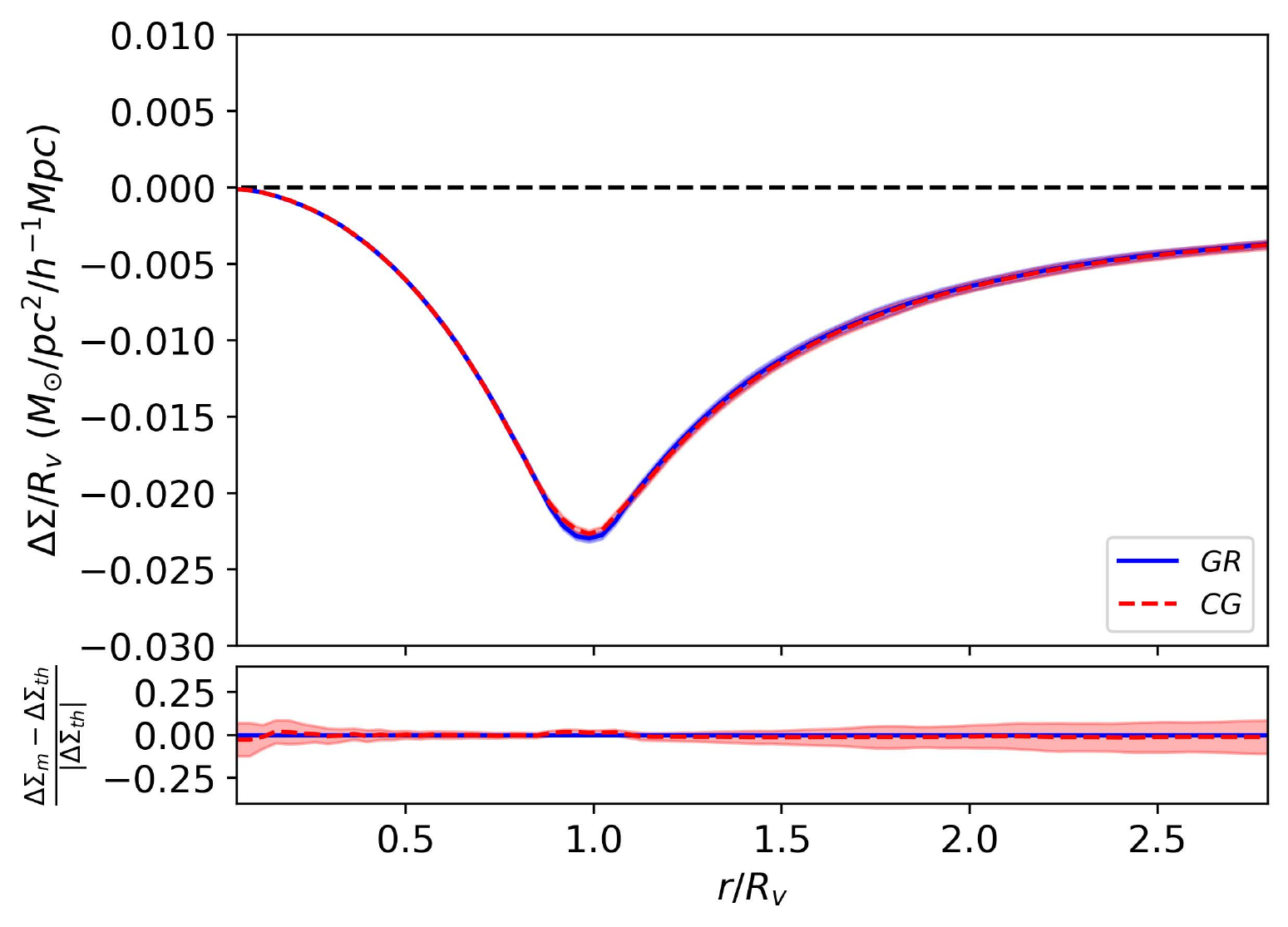}}
\caption{Lensing signals calculated in different models. For both panels, the blue lines represent the calculation results of GR model, and the red lines represent CG model. {The shadow regions represent the error propagated from the jackknife error in void profiles}. \textit{Left}: Impact of the void density profile on the lensing signal in the GR and CG gravity models. Lensing signals were calculated using the same deflection angle (i.e., the same Poisson equation) but different density profiles. \textit{Right}: Impact of the deflection angle on the lensing signal in two theories. This calculation used the same void density profile but different Poisson equations for the two gravity models. It can be seen that the discrepancies between these two models is negligible compared with that in left panel, i.e., from the modification on void profile, illustrating that no significant effect from the modification of the Poisson equation. }
\label{dsigr}
\end{figure*}

We calculate the ESD based on the GR and CG gravity models described in Sec.~\ref{lencal}. {During the calculation, we use the open-source package scipy.fft.fht\citep{2020SciPy-NMeth} to do the Fourier transform and its inverse.} Effects from the change of the void profile and the deflection angle are separately considered. First, the effects of different void profiles are investigated. In the left panel of Fig.~\ref{dsigr}, we showcase the lensing signals around voids in GR and CG simulations, while the lensing signal is calculated with the same deflection angle predicted in GR. {The errorbar in the figure is estimated in this way: The jackknife error of the lensing signal is estimated in each single simulation, then we combine these signals and averaged them to obtain the mean signal while propagating the jackknife error in each realization to the final signal.} As shown in the figure, the lensing signal of CG theory is larger than that of GR {(even with a relatively large errorbar) which is consistent with the properties in void profile (Fig.~\ref{void_prof}), where CG voids have an emptier profile.} This can be explained by the fifth force, which provides an additional contribution to enhancing the structure formation. Second, the lensing signals are calculated on the same input void profile (from GR), but with different deflection angles adapted for the two gravity models, respectively. The error bar is estimated in the same way as that used in the case which only modifies the dark matter field. Our results are shown in the right panel of Fig.~\ref{dsigr}, showing that the influence of the change in the deflection angle is negligible compared to that of the different density profiles. This is expected since $G_{\text{eff}}$ differs from $G$ only on the order of thousandth (see Fig.~5 in \citet{PhysRevD.102.043510}).

Based on these calculations, we can conclude that CG theory affects the lensing signal mostly by changing the void profile rather than changing the deflection angle of light, though to a fairly limited extent. Also, it is expected that in the following simulation results, if one only works on a modified deflection angle, there should be a less distinguishable discrepancy. 

\subsection{Lensing signal from simulation} \label{vlsignal}
We measure the lensing statistics $\Delta\Sigma$ (ESD) from our ray-tracing simulations. Based on Eqs.~\ref{gt_kappa} and \ref{esd_kappa}, ESD can be related to tangential shear $\gamma_t$ by:
\begin{equation}
    \Delta\Sigma=\Sigma_{\text{crit}}\gamma_{t}, \label{esd_gt}
\end{equation}
where $\Sigma_{\text{crit}}$ is defined as:
\begin{equation}
    \Sigma_{\text{crit}}=\frac{1}{4{\pi}G}\frac{D_A(z_{\text{s}})}{D_A(z_{\text{L}})D_A(z_{\text{L}},z_{\text{s}})},
\end{equation}
where $D_A$ is the angular diameter distance. The estimator of the ESD:
\begin{equation}
    \widehat{\Delta\Sigma}
    =\frac{\sum_{i,j}{w_{ij}e_{t,ij}\Sigma_{\text{crit},ij}}}{\sum_{ij}{w_{ij}}}, \label{esti_esd}
\end{equation}
where $e_t$ represents the ellipticity of the background galaxy, $w$ represents the weight of each source-lens pair, and the summation is done over all source-lens pairs.

We randomly sample 5,582,821 background galaxies (corresponding to a number density of $10.00~\text{arcmin}^{-2}$) on a source plane located at $z=0.75$, rendering their projection positions to satisfy a uniform distribution. And further, each background galaxy is assigned with a redshift dispersion of ${\Delta}z=0.001$. In practice, the shear components we concern in the ray-tracing simulation are the ellipticities of background galaxies, while for foreground voids, we select those in slices that constitute the lens plane at $z=0.3$, of which the redshifts are determined by their comoving distance to us along the line of sight, but also with a dispersion of ${\Delta}z=0.001$. We employ \textsc{swot} \citep{2012A&A...542A...5C} to measure the ESD around foreground voids. 

\begin{figure*}
    \centering
    \subfigure{
    \includegraphics[scale=0.5]{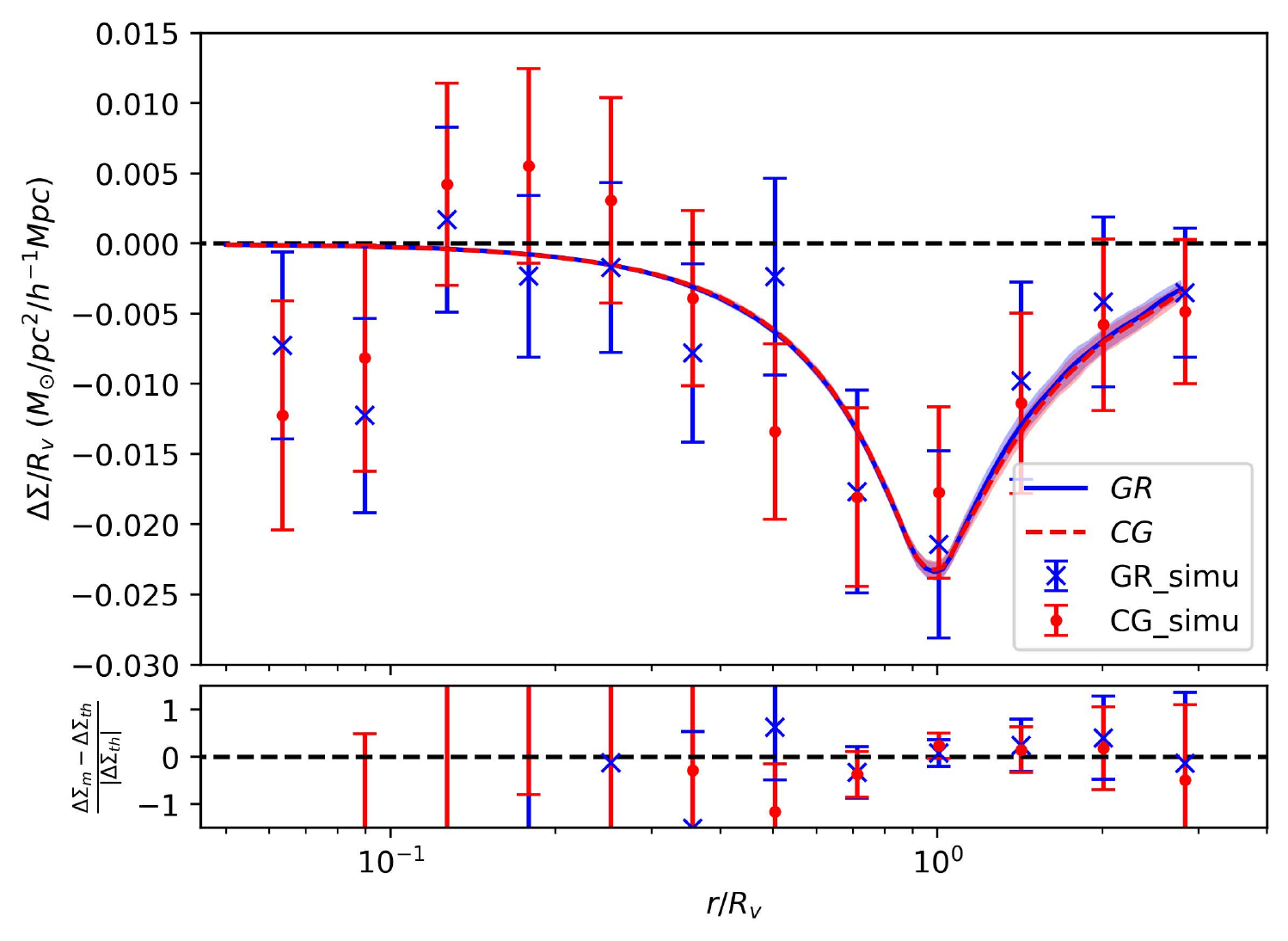}}\subfigure{
    \includegraphics[scale=0.5]{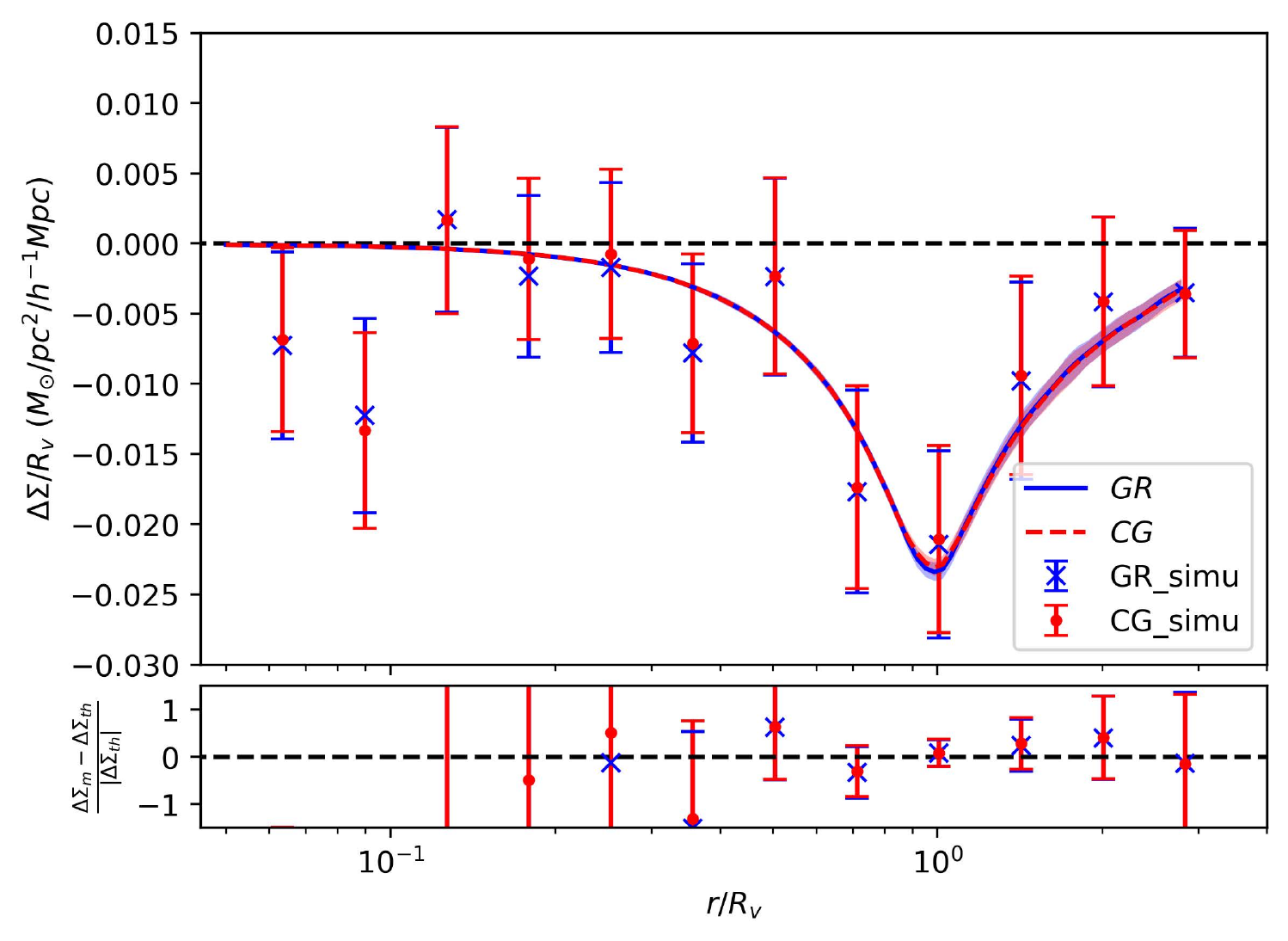}}
    \caption{ESD measured in simulation (dots with error bar) versus theoretical predictions (lines) from GR and CG. The error bars in both of the theories and simulations are obtained from the jackknife method. \textit{Left}: Only the void density profile is changed, representing the effects on the structure formation. The red color represents CG model and the black represents CG. It can be seen that the two models are consistent with the corresponding simulation results nevertheless with the large error bar. \textit{Right}: Only the deflection angle is changed. The void profile is obtained from GR simulation. In this subfigure, we only consider an extreme case. It can be seen that consistent with the findings in theoretical predictions, the simulation results of the two models are indistinguishable.}
    \label{esd_mvst}
\end{figure*}

Our results for GR and CG models in comparison are displayed in Fig.~\ref{esd_mvst}. The left panel is dedicated to the measurement on different dark matter fields but with the same deflection angle, whilst vice versa for the right panel. From the left panel, we see the lensing signal of GR is systematically larger than that of CG but the difference is not significant. While for the right panel, an expected result shows up as no significant deviation for GR from the CG simulation. 

In Sec.~\ref{choose} we discussed lensing statistics $\Delta\Sigma$ and the equivalence between this and physical excess surface mass density. Since we find that the change to the lensing signal from a varying deflection angle is incomparable to that from a differing void density profile, it is still valid to interpret the differential projected mass density as the lensing signal here. The discussions hereafter will be limited to the case of varying dark matter density fields alone. Nevertheless, in other MG theories, it may not be such a simple case. That is, our method should be more useful in studying those MG theories which can greatly modify the deflection angle. 

\begin{figure*}
    \centering
    \includegraphics[scale=0.5]{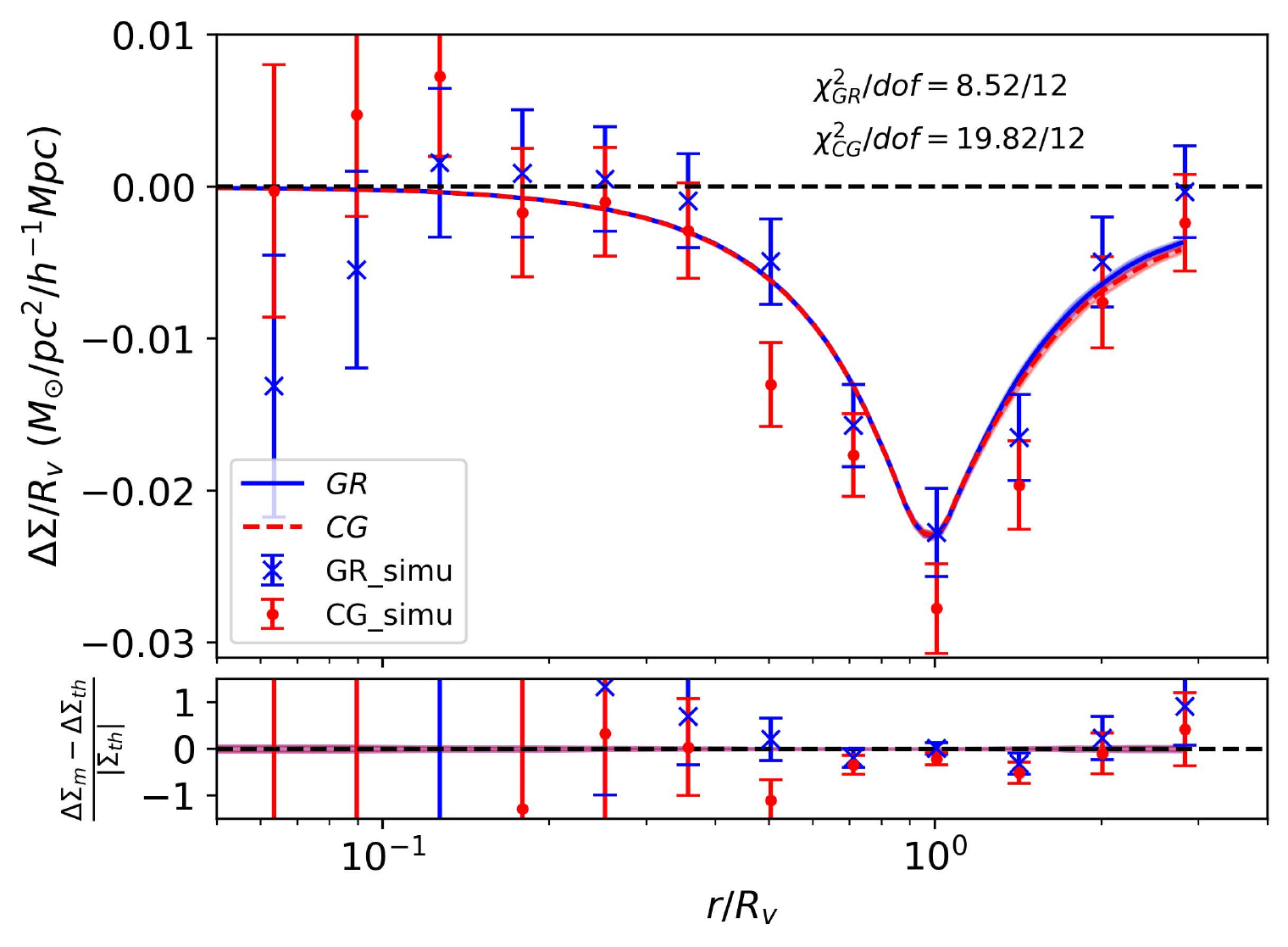}
    \caption{The same as Fig.~\ref{esd_mvst}, but estimated by averaging measurements from 9 lightcones. Besides we also show the deviation between simulation and theory on the bottom panel. For theoretical predictions, we calculate the ESD using stacked void profiles in 9 lightcones, respectively, and combine these 9 results as the final predictions. The jackknife error in each single measurement is combined as the error bar.}
    \label{vlens_full}
\end{figure*}

We are interested in the capability of void lensing statistics in distinguishing different gravity models. The first thing is to reduce the statistical uncertainty. To accomplish this, we employ the same procedure to generate 9 background galaxy catalogs and 9 corresponding foreground void catalogs from 9 lightcones. By averaging the resulting lensing signals, we are able to decrease the statistical uncertainty by a factor of $1/3$ relative to a single realization. Our simulation results then represent an observation case with an effective survey area of $\sim{1395~\rm deg^2}$. In order to account for shape noise, we incorporate noisy shear components $(e_1, e_2)$ into our measurements, where $e=(\gamma+n)/(1+n\gamma^*)$. Here we use the convention of $e=e_1+{\rm i}e_2$, $\gamma=\gamma_1+{\rm i}\gamma_2$ (shear from simulation) and $n=n_1+{\rm i}n_2$ (random elliticity). The random ellipticity components are drawn from Gaussian distribution with zeros mean and a standard deviation of $\sigma_e=0.3$. The results are established in Fig.~\ref{vlens_full}.

Clearly, in Fig.~\ref{vlens_full} the statistical uncertainties for both GR and CG models are significantly reduced. This allows us to assess the capability of void lensing as a probe to detect the discrepancy between GR and CG gravity. We utilize the GR and CG simulation data separately as the observed signal for a fitting of theoretical predictions. The goodness of the fitting is manifested in the $p$-value. In each simulation, the number of data points is 12, of which the number acts as the degree of freedom given no additional free parameters from the two gravity models.

We first assume that the observed signals are consistent with those in GR simulation, and then calculate the $\chi^2$ value of the two theoretical predictions. The $\chi^2$ value of the GR model is 8.52, indicating that the model can be rejected at ${0.33\sigma}$: ${p_{\text{GR}}=0.744}$. In contrast, the $\chi^2$ value of the CG model is 8.62, corresponding to a $p$-value of 0.735, which implies that the CG model can be rejected at ${0.34\sigma}$. Therefore we cannot reject the CG model although the GR model has a smaller $\chi^2$ value than the CG model. 

We then assume the CG simulation represents as the real observation and repeat the analysis. The $\chi^2$ of GR model is 20.537, corresponding to a $p$-value of 0.058. This implies that GR gravity is rejected by the CG simulation at ${1.90\sigma}$. In comparison, the $\chi^2$ of CG model is 19.823, which is rejected at ${1.81\sigma}$ with a $p$-value of 0.071. The results present a preference for CG gravity in the CG simulation, but it is not significant because the differences between CG and GR models are much smaller than those between the simulations (see Fig~\ref{vlens_full}). Furthermore, the consistency between CG theory and simulation is less than expected, which could be due to the simplification of the theoretical model, incomplete estimation of cosmic variance, or other unknown systematics.

\subsection{Forecast for the next generation survey}

\begin{figure*}
    \centering
    \includegraphics[scale=0.5]{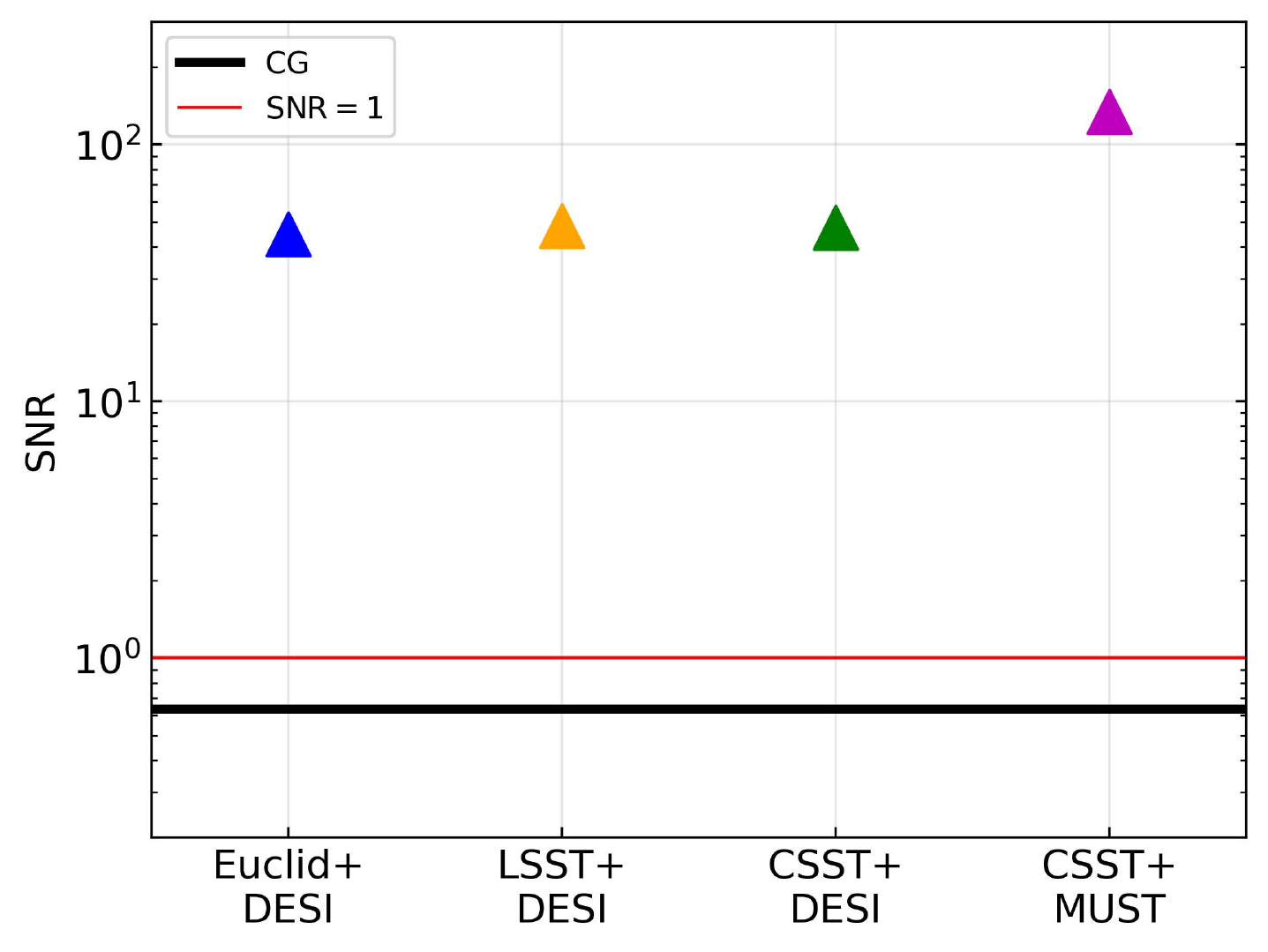}
    \caption{Comparison of Signal-to-Noise Ratio (SNR) of the differences between CG and GR in ESD predicted by different future survey conditions. The black solid line represents SNR in our simulation. The colored triangles represent the SNR of differences between GR and CG. Different colors represent different combinations of spectroscopic and photometric surveys.}
    \label{forefig}
\end{figure*}

In the previous subsection, we conclude that the two models are not distinguishable in the case of our simulations. It is interesting to consider whether the next-generation survey will be able to detect differences between those two models. We can make a simple prediction for the probability of distinguishing these two models with the help of the Signal-to-Noise Ratio (SNR), which is defined as:

\begin{equation}
    {SNR=\sum_{i}{\frac{\Delta\Sigma_{\rm CG}-\Delta\Sigma_{\rm GR}}{\sqrt{(\sigma_{i,\rm GR}^2+\sigma_{i,\rm CG}^2)}}}}, \label{snr}
\end{equation}
where $\Delta\Sigma_{i}$ is the void lensing signal and ${\sigma^2}$ represents the errors. Here for the CG model, only the dark matter field is modified. It should be also noticed that we ignore the covariance between data points which will result in an overestimation of the SNR. The solid black line in Fig.~\ref{forefig} represents the SNR of our simulation, whose value is ${\sim0.6}$, indicating that it might be impossible to distinguish these two gravity models (which is consistent with the results in Sec.~\ref{vlsignal}). Then we investigate how the next-generation weak lensing survey can raise the value of SNR. There are two possible approaches to improving the SNR. The first is to increase the void number density, requiring expanding the spectroscopic survey. In this forecast, we consider two spectroscopic surveys: DESI \citep{2016arXiv161100036D}, which has a void number density of ${\sim1~\rm arcmin^{-2}}$ (we assume that the number of voids is proportional to the number of galaxies, so if the galaxy number density is 10 times greater than in our simulations, the void number density is estimated to be 10 times greater than ours); and the Multiplexed Survey Telescope (MUST)\footnote{\url{https://must.astro.tsinghua.edu.cn/must/index.html}}, which has an optimally evaluated galaxy number density that is 8 times larger than that of DESI, resulting in a corresponding void number density of ${8~\rm arcmin^{-2}}$. The second approach to improve the SNR involves increasing the background galaxy number and widening the survey area, which can be achieved by future photometric surveys. Three next-generation survey missions are taken into account: Euclid, LSST and CSST. Euclid is expected to cover a survey area of ${15,000~\rm deg^2}$ with a background galaxy number density of ${\sim30.0~\rm arcmin^{-2}}$ \citep{2011arXiv1110.3193L}. LSST, on the other hand, aims to scan ${20,000~\rm deg^2}$ with a galaxy number density of ${26~\rm arcmin^{-2}}$ \citep{2013MNRAS.434.2121C}. Finally, the CSST survey has a total survey area of ${17,500~\rm deg^2}$ with a galaxy density of ${28.89~\rm arcmin^{-2}}$ \citep{2019ApJ...883..203G}. Based on the void number density in next-generation spectroscopic surveys, the galaxy number density, and the survey area in the next-generation photometric survey, we present the improved SNR values in Fig.~\ref{forefig}. The labels on the bottom axis of the figure are formatted as ``spec survey''${+}$``photo survey'' where, for example ``Euclid${+}$DESI'' represents the SNR with a void number density of DESI, a survey area and the background galaxy density of Euclid. As illustrated in the figure, all four cases can increase the SNR value to ${\sim40-50}$. This suggests that it may be possible to distinguish the CG and GR models in the next-generation survey data. {However, it is important to note that our forecast is a simple case and does not consider the dependency of the galaxy bias on the local density. As pointed out in \citet{2017MNRAS.469..787P} that the galaxy bias is different in the voids we considered in this paper. Therefore, further efforts are needed to obtain a more accurate evaluation of the possibility of distinguishing these two models by void lensing.}

Besides, we also forecast that if the effects w.r.t. the deflection angle can be detected in the next-generation surveys. For this case, the SNR in our simulation is lower than ${10^{-4}}$, and the next-generation survey can only increase the value to about 2 orders of magnitude, which is still far from unity. Therefore, it seems that for CG gravity studied in this paper, it is unnecessary to run a ray-tracing simulation to obtain the lensing signals. However, we also point out that this conclusion is only valid for the CG gravity considered in our study, and for other gravity models especially for which the deflection angle is greatly changed, it is necessary to run a ray-tracing simulation to check the correction of the lensing signals obtained from the N-body simulation.

\section{Summary and Conclusion} \label{sec:sum&conclu}
In this study, we focused on investigating the weak lensing signal around cosmic voids in the context of CG gravity, and further tested the reliability to neglect the effects from the deflection angle in this model. Instead of directly projecting the dark matter density field along the line of sight in N-body simulation, we utilized a ray-tracing technique to generate convergence and shear maps on the background at redshift $z=0.75$. The foreground mock galaxies at $z=0.3$ generated by SHAM and voids are identified from them using {\footnotesize DIVE} method, and the background galaxies are randomly sampled from a uniform distribution on a single source plane. The effective survey area in our simulation (${\sim1395~\rm deg^2}$) as well as the number of background galaxies (${10~\rm arcmin^{-2}}$) are chosen so as to match the current KiDS-\textit{like} weak lensing survey. The lensing signals are measured from the simulations, which constitute the `real observations' confronted by theoretical predictions. We conclude that:

(1) Regarding the void lensing signal, the effect of the CG gravity on the deflection angle is incomparable to that on the void profile. Owing to a less distinguishable $G_{\text{eff}}$ from the gravitational constant, the interaction between particles and particles (photons) in CG is very similar to that in GR. Nonetheless, the cumulative effect on structure formation should still be observable in our simulations spanning from $z=49$ to $z=0$. In contrast, the effect on the deflection angle is transient, even when accounting for foreground matter between $z=0$ and $z\sim1$, which determines its subordinate status. However, some other studies such as \citet{PhysRevD.98.023511, Barreira_2015_08}
hold the opposite view, suggesting that the void lensing signals in CG can be approximately twice as strong as GR due to the change to the lensing potential (i.e., the deflection angle). This difference can be explained from two aspects: a) the inclusion of a potential term as a source to drive late acceleration in the Lagrangian of our CG model, which is different from previous literature; and b) the prevalent tracker ansatz proposed by \citet{PhysRevLett.105.111301} is no longer suitable for our model due to the participation of the potential term. To this end, we employ numerical techniques to solve the autonomous equations while preserving high-order terms in the field equations that were dropped in \citet{Barreira_2015_08}.

(2) For both GR and CG, the measured signals are consistent with the theoretical predictions. Since the lensing signal is barely sensitive to the change of deflection angle, the detection of $\Delta\Sigma$ can be approximately transferred to a measurement of foreground void density profile. The theoretical prediction, which is a projection of the 3D density profile, is expected to agree well with the measurement from our ray-tracing simulations. The slight overestimation seems to be spurious and can be attributed to the small boxsize of the simulation, the number of realizations, and the simplification of the theoretical model. Despite this, our methodology has two major advantages: a) It can be used as a test for lensing studies in other MG theories, particularly those with significantly changed deflection angles. This test is crucial to verify eq. \ref{esd_kappa}, i.e., ESD is an accurate estimation of lensing potential. b) The use of ray-tracing techniques to measure the lensing signal from N-body simulations is more reliable than a direct projection to the mass density, as the latter lacks the change of the deflection angle and the total distribution of mass between the source and lens.

(3) The two gravity models are indistinguishable in our work, where the configuration of our simulations is up to an effective survey area of ${\sim 1395~\deg^2}$ and a galaxy number density of $10~\text{arcmin}^{-2}$. It is possible to distinguish these two models with the help of the extended survey area and enlarged galaxy number density. By analyzing the SNR of the differences between the two gravity models, we find that in the next-generation surveys, the SNR reaches from ${\sim0.6}$ to several decades. However, when it comes to data processing, many other difficulties such as systematic calibrations must be overcome. {Meanwhile, the dependence of the galaxy bias on the local density is not considered in the forecast and needs to be modeled more accurately in future work.} Additionally, even considering the next-generation survey, the effect due to the modification of the deflection angle is also impossible to be detected. In other words, in the Stage-IV era, it is still valid to make the approximation that regarding the lensing statistics ESD as the projection of dark matter field in CG gravity. This is expected since the effective gravitational constant ${G_{\text{eff}}}$ is very similar to the Newtonian gravitational constant ${G}$ in this gravity. We also would like to stress that when studying the lensing effect in the gravity models, especially those in which changes in the deflection angle have a larger influence on the lensing signal than changes in the dark matter field, it is necessary to run a ray-tracing simulation to obtain a more reliable lensing signal.

As a bonus scene in the final, we would like to stress the feasibility of measuring the lensing signal from N-body simulations via a projected 3D density profile for CG gravity, after comparison to the use of the ray-tracing technique. This conclusion should be valid in the context of a CG gravity that can appreciably modify the DM density field other than the deflection angle. It is a promising algorithm to study lensing effects in those gravity models that apparently modify the path of the photons. A practical test of our methodology on real data-those for example, from LSST, Euclid, CSST and etc. will be of more interest and is left for future works, along with sophisticated modeling of observational systematics. Additionally, exploring how to distinguish CG gravity from GR through a synergy of void lensing with other observables (such as galaxy clustering and galaxy-galaxy lensing), as well as determining the extent to which we can constrain the model, are both worth studying.

\begin{acknowledgements}

We thank the anonymous reviewer for the valuable comments and suggestions to improve our paper. We would like to thank Bikash R. Dinda, Md. Wali Hossain and Anjan A. Sen for valuable discussions. CS would also like to thank Ningchen Bai for discussions on the theory of Modified Gravity. We acknowledge the support from the science research grants from the China Manned Space Project with No. CMS-CSST-2021-A01, No. CMS-CSST-2021-A03, No. CMS-CSST-2021-A07 and No. CMS-CSST-2021-B01. HYS acknowledges the support from NSFC of China under grant 11973070, Key Research Program of Frontier Sciences, CAS, Grant No. ZDBS-LY-7013 and Program of Shanghai Academic/Technology Research Leader. JY, CZ acknowledges the support from the SNF 200020\_175751 and 200020\_207379 ``Cosmology with 3D Maps of the Universe'' research grant. QW acknowledges the support from National Natural Science Foundation of China (Grant No. 11988101). XKL, AZ acknowledges the support from NSFC of China under Grant No. 12173033. L.X. is supported by the National Research Foundation of Korea (NRF) through Grant No. 2020R1A2C1005655 funded by the Korean Ministry of Education, Science and Technology (MoEST), and by the faculty research fund of Sejong University. This work made use of the High Performance Computing Resource in the Core Facility for Advanced Research Computing at Shanghai Astronomical Observatory.
\end{acknowledgements}

\bibliography{void_lensing}{}
\bibliographystyle{aasjournal}



\end{document}